\documentclass[preprint, 2pt]{aastex}
\usepackage{graphicx,natbib}
\usepackage{xcolor}
\usepackage{url}
\usepackage{amsmath}

\begin{document}
\title{A Stochastically Forced Time Delay Solar Dynamo Model: Self-Consistent Recovery from a Maunder-like Grand Minimum Necessitates a Mean-Field Alpha Effect}
\author{Soumitra Hazra\altaffilmark{1}, D\'ario Passos\altaffilmark{2,3,4} and Dibyendu Nandy\altaffilmark{1,5} }
\affil{$^1$Department of Physical Sciences, Indian Institute of Science Education and Research, Kolkata}
\email{s.hazra@iiserkol.ac.in}
 \affil{$^2$CENTRA-IST, Instituto Superior
T\'ecnico, Av. Rovisco Pais, 1049-001 Lisboa, Portugal}
\email{dariopassos@ist.utl.pt}
\affil{$^3$Departamento de F\'\i sica, Universidade de \'Evora, Col\'egio Ant\'onio Luis Verney, 7002-554 \'Evora, Portugal}
\affil{$^4$ D\'epartment de Physique, Universit\'e de Montr\'eal, C.P. 6128 Centre-ville, Montr\'eal, Qc, Canada  H3C-3J7 }
\affil{$^5$Center of Excellence in Space Sciences India, IISER Kolkata, Mohanpur 741252, West Bengal, India}
\email{dnandi@iiserkol.ac.in}
\begin{abstract}

Fluctuations in the Sun's magnetic activity, including episodes of grand minima such as the Maunder minimum have important consequences for space and planetary environments. However, the underlying dynamics of such extreme fluctuations remain ill-understood. Here we use a novel mathematical model based on stochastically forced, non-linear delay differential equations to study solar cycle fluctuations, in which, time delays capture the physics of magnetic flux transport between spatially segregated dynamo source regions in the solar interior. Using this model we explicitly demonstrate that the Babcock-Leighton poloidal field source based on dispersal of tilted bipolar sunspot flux, alone, can not recover the sunspot cycle from a grand minimum. We find that an additional poloidal field source effective on weak fields -- e.g., the mean-field $\alpha$-effect driven by helical turbulence -- is necessary for self-consistent recovery of the sunspot cycle from grand minima episodes.

\end{abstract}

\maketitle
\section{Introduction}

Sunspots, which are strongly magnetized regions, play a key role in governing the activity of the Sun. The number of sunspots observed on the solar surface waxes and wanes with time generating the 11-year solar cycle. While there is a small variation in this periodicity, fluctuations in the amplitude of the solar cycle are large. Extreme fluctuations are manifest in grand maxima episodes -- when the cycle amplitudes are much higher than normal, and grand minima episodes -- when the cycle amplitudes fall drastically, even leading to the disappearance of sunspots for an extended period of time. The most striking evidence of such a minimum in the recorded history of sunspot numbers is the so-called Maunder minimum between 1645 and 1715 AD \citep {Eddy88}. The lack of sunspots during this period is statistically well-proven and is not due to the lack of observations -- which covered 68\% of the days during this period  \citep{hoyt96}. The occurrence of these solar activity extremes is correlated with temperature records over millennium scale \citep{Usoskin2005}; the solar Maunder minimum coincided with the severest part of the Little Ice Age -- a period of global cooling on Earth.

Over the last decade, solar activity reconstructions based on cosmogenic isotopes and geomagnetic data \citep{Usoskin2000, Usoskin2003, Miyahara2004,Usoskin2007,Steinhilber2010,Lockwood2011}, which are indirect proxies for probing long-term solar activity have brought to the fore various properties of these grand minima episodes. These observations show that there have been many such activity minima in the past; however the solar cycle has recovered every time and regained normal activity levels. There is some evidence for persistent, but very weak amplitude cycles during the Maunder minimum and a slow strengthening of cycle amplitudes to normal levels during the recovery phase. While the general perception was that the onset of the Maunder minimum was sudden, a recent reconstruction based on historical sunspot records has challenged that notion indicating that the onset phase of the minimum may have been gradual \citep{Vaquero2011}.

A magnetohydrodynamic (MHD) dynamo mechanism, involving interactions of plasma flows and magnetic fields drives the solar cycle. Our understanding of the solar dynamo, see e.g., the reviews by  Ossendrijver \citep{Ossendrijver2003} and Charbonneau \citep{Charbonneau2010}, is based on the generation and recycling of the toroidal and poloidal components of the Sun's magnetic field. The toroidal magnetic field is produced by stretching of poloidal field lines by differential rotation -- a process termed as the $\Omega$-effect \citep{Parker1955}. It is thought this process is concentrated near the base of the solar convection zone (SCZ) -- where the upper part of the tachocline (a region of strong radial gradient in the rotation) and overshoot layer (which is stable to convection) offers an ideal location for toroidal field amplification and storage. Sufficiently strong toroidal flux tubes are magnetically buoyant and erupt radially outwards producing sunspots where they intersect the solar surface.

For the dynamo to function, the poloidal component has to be regenerated back from the toroidal component, a step for which, diverse propositions exist. The first such proposition invoked helical turbulent convection as a means of twisting rising toroidal flux tubes to regenerate the poloidal component (a mechanism traditionally known as the the mean field $\alpha$-effect \citep{Parker1955}). Numerous dynamo models based on the mean-field $\alpha$-effect were constructed and such models enjoyed a long run as the leading contender for explaining the origin of the solar cycle \citep{Charbonneau2010}. However, subsequent simulations of the dynamics of buoyant toroidal flux tubes and observational constraints set by the tilt angle distribution of sunspots pointed out that the toroidal magnetic field at the base of the SCZ must be as high as $10^5$ G \citep{silva93,fan93,Caligari95}; such strong toroidal flux tubes being one order of magnitude stronger than the equipartition magnetic field in the SCZ would render the mean field $\alpha$-effect ineffective. This consideration revived interest in an alternative mechanism of poloidal field production based on the flux transport mediated decay and dispersal of tilted bipolar sunspots pairs in the near-surface layers \citep{Babcock61, Leighton69}, hereby, referred to as the Babcock-Leighton mechanism. \\

In the last couple of decades, multiple dynamo models have been based on this idea \citep{Durney97, Dikpati99, Nandychoudhuri02, Chatterjee04, Munoz09} and have successfully reproduced many nuances of the solar cycle. Some \citep{Tobias06,Bushby07,Cattaneo09} have criticised the usage of such mean-field dynamo models to predict the solar cycle, however it should be noted that recent studies \citep{Simard2013, Dube13} indicate that if input profiles are extracted from three-dimensional full MHD simulations and fed into two-dimensional mean-field dynamo models, they are capable of producing qualitatively similar solutions to those found in the full MHD simulations. The major advantage of the mean-field dynamo framework is that it allows for much faster integration times compared to the full MHD simulations and are therefore computationally efficient as well as physically transparent. Recent observations also lend strong support to the Babcock-Leighton mechanism \citep{Dasi-Espuig2010,Munoz13} and this is now believed to be the dominant source for the Sun's poloidal field. Surface transport models \citep{Wang89, van98} also provide theoretical evidence that this mechanism is in fact operating in the solar surface. Randomness or stochastic fluctuations in the Babcock-Leighton poloidal field generation mechanism is an established method for exploring variability in solar cycle amplitudes \citep{Charbonneau2000, Charbonneau04,Charbonneau05,PassosLopes2011,Passos2012,Choudhuri2012} as are deterministic or non-linear feedback mechanisms \citep{Wilmot-Smith05,Jouve2010}. Stochastic fluctuations within the dynamo framework are physically motivated from the random buffeting that a rising magnetic flux tube endures during its ascent through the turbulent convection zone and from the observed scatter around the mean (Joy's law) distribution of tilt angles. It is to be noted that similar fluctuations are to be expected in the mean-field $\alpha$ effect as well \citep{hoyng88} and such phenomenon can be explored within the framework of truncated mean-field dynamo models \citep{yoshimura75}.

Since the two source layers for toroidal field generation (the $\Omega$ effect) and poloidal field regeneration (the $\alpha$-effect) are spatially segregated in the SCZ, there must be effective communication to complete the dynamo loop.  Magnetic buoyancy efficiently transports toroidal field from the bottom of the convection zone to the solar surface. On the other hand, meridional circulation, turbulent diffusion and turbulent pumping share the role of transporting the poloidal flux from the surface back to the solar interior \citep{Karak2012} where the toroidal field of the next cycle is generated thus keeping the cycles going. Thus, there is a time delay built into the system due to the finite time required for transporting magnetic fluxes from one source region to another within the SCZ.

Based on delay differential equations and the introduction of randomness on the poloidal field source, here we construct a novel, stochastically forced, non-linear time delay dynamo model for the solar cycle to explore long-term solar activity variations. We particularly focus our investigations on the recovery from grand minima phases and demonstrate that the Babcock-Leighton mechanism alone -- which is believed to be the dominant source for the poloidal field -- cannot restart the solar cycle once it settles into a prolonged grand minimum. The presence of an additional poloidal field source capable of working on weak magnetic fields, such as the mean field $\alpha$-effect is necessary for recovering the solar cycle.

\section{Stochastically Forced, Non-Linear, Time Delay Solar Dynamo Model}

The model is an extension of the low order time delay dynamo equations explored in an earlier study involving one of us \citep{Wilmot-Smith06}. This model was derived considering only the source and dissipative mechanisms in the dynamo process. All space dependent terms were removed and instead the physical effect of flux transport through space was captured through the explicit introduction of time delays in the system of equations.

The time delay dynamo equations are given by
 \begin{eqnarray}
    \frac{dB_\phi(t)}{dt}&=& \frac{\omega}{L} A(t-T_0)-\frac{B_\phi(t)}{\tau}\ \\
    \frac{dA(t)}{dt}&=&\alpha_0 {f_1} (B_\phi(t-T_1)) B_\phi(t-T_1) -\frac{A(t)}{\tau} \, ,
 \end{eqnarray}
where $B_\phi$ represents toroidal field strength and A represents poloidal field strength. The evolution of each magnetic component is due to the interplay of the source and dissipative terms in the system. In the toroidal field evolution equation, $\omega$ is the difference in rotation rate across the SCZ and $L$ is the length of SCZ. The dissipative term is governed by turbulent diffusion, characterized by the diffusion time scale ($\tau$). The parameter $T_0$ is the time delay for the conversion of poloidal field into toroidal field and is justified by the finite time that the meridional circulation or turbulent pumping takes to transport the poloidal magnetic flux from the surface layers to the tachocline. $T_1$ is the time delay for the conversion of toroidal field into poloidal field and accounts for the buoyant rise time of toroidal flux tubes through the SCZ. The meridional circulation timescale is about 10 yr for a peak flow speed 20 m$s^{-1}$ (from mid-latitudes at near-surface layers  to mid-latitudes above the convection zone base; see \cite{Yeates2008} for detailed calculation of meridional circulation time scale). Another dominant flux transport mechanism for downward transport of magnetic field could be turbulent flux pumping with a timescale of about one yr (with a relatively high pumping speed of 5 m$s^{-1}$). The buoyant rise time of flux tubes from the SCZ base to surface is about three months (assuming the rise timescale is of the order of Alfvenic time scale, which is also a general agreement with simulations; see also \cite{fan93}). As the magnetic buoyancy time scale is much shorter compared to the meridional circulation (or turbulent diffusion or flux pumping) timescale, we assume $T_1<<T_0$. Since it is not clear which is the most dominant flux transport mechanism - meridional circulation or turbulent pumping, we explore our model in two different regimes of operation to test for robustness. In one setup, we consider $T_0 = 4 T_1$ (if $T_0$ corresponds to pumping time scale) and $T_0= 40 T_1$ (if $T_0$ corresponds to meridional circulation time scale). This model setup mimics spatial separation between two source layers in the Sun's convection zone and the role of magnetic flux transport between them and is therefore physically motivated. On the other hand, due to its nature, this model is amenable to long time-integration without being computationally expensive.

\begin{figure}[t!]
 \centering
\includegraphics*[width=\linewidth]{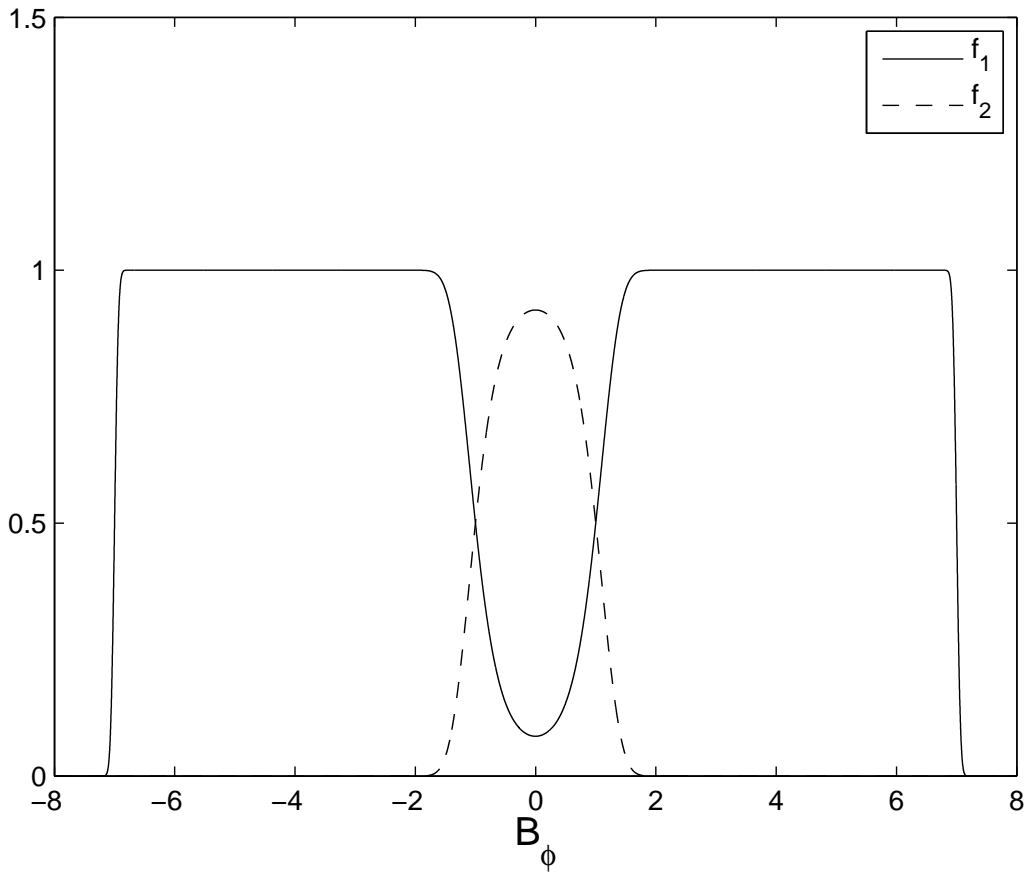}
  \caption{Profile of the quenching function $f_1$ for the Babcock-Leighton $\alpha$ and $f_2$ for the weak, mean field $\alpha$-effect (described later in the text). The plot of $f_1$ corresponds to parameters $B_{min}=1$ and $B_{max} =7$ and $f_2$ corresponds to $B_{eq}=1$ (all in arbitrary code units).}
  \label{fig1}
\end{figure}

To account for quenching of the Babcock-Leighton poloidal source $\alpha$, we take a general form of $\alpha$, i.e $\alpha=\alpha_0 {f_1}$, where $\alpha_0$ is the amplitude of the $\alpha$ effect and $f_1$ is the quenching factor approximated here by a nonlinear function
\begin{eqnarray}
     f_1=\frac{[1+erf(B^2_{\phi}(t-T_1)-B^2_{min})]}{2}\\ \nonumber
     \cdot \frac{[1-erf(B^2_{\phi}(t-T_1)-B^2_{max})]}{2}.
\end{eqnarray}
Figure \ref{fig1} depicts this quenching function, constructed with the motivation that only flux tubes with field strength above $B_{min}$  (and not below) can buoyantly rise up to the solar surface and contribute to the Babcock-Leighton poloidal field source, i.e., sunspots \citep{Parker1955} and that flux tubes stronger than $B_{max}$ erupt without any tilt therefore quenching the poloidal source \citep{silva93,fan93}. Accounting for these lower and upper operating thresholds for the Babcock-Leighton poloidal source is fundamentally important for the dynamics.

\begin{figure}[t!]
 \centering
\includegraphics*[width=\linewidth]{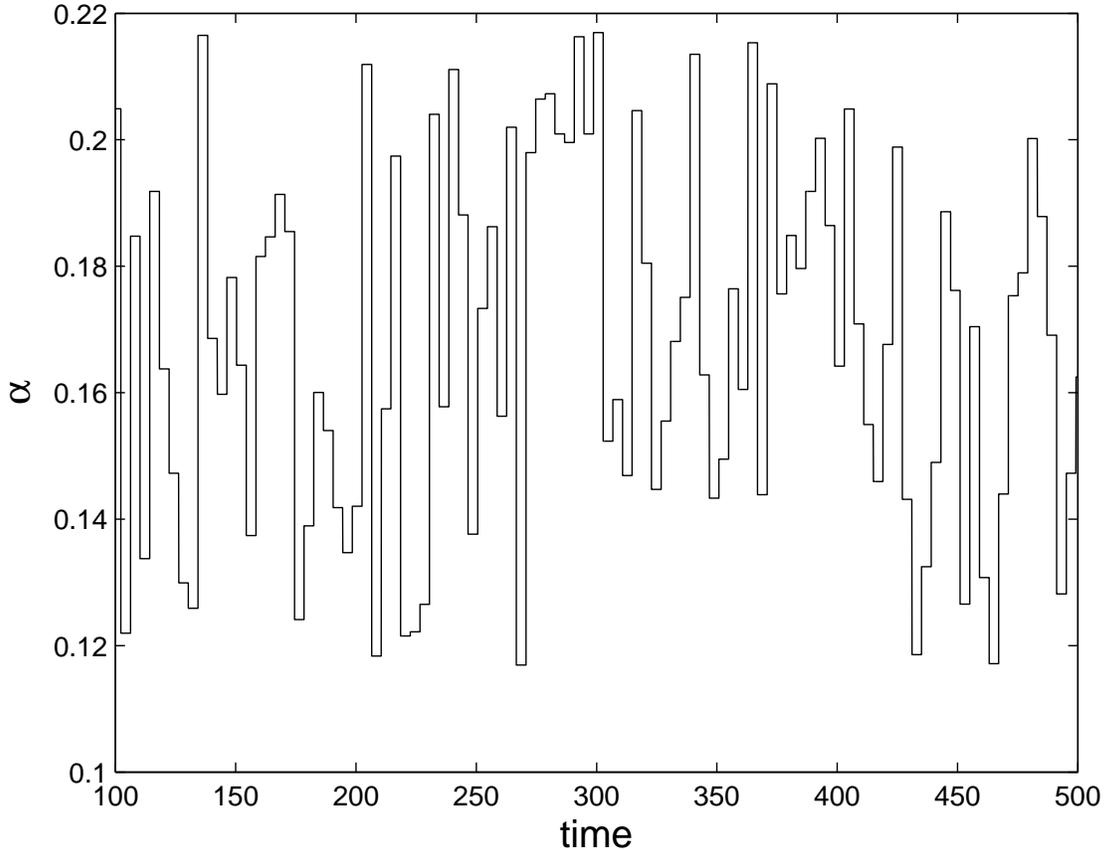}
  \caption{Stochastic fluctuations in time in the poloidal source term $\alpha$ at a level of 30\% ($\delta=30$) using our random number generating programme.}
\label{fig2}
\end{figure}

Our aim here is to explore the impact of stochastic fluctuations in this time delay solar dynamo model. For $\alpha=\alpha_0$, we get a strictly periodic solution. In order to introduce stochastic fluctuations, we redefine $\alpha$ as
\begin{equation}
    \alpha=\alpha_0\,[1+\frac{\delta}{100} \sigma(t,\tau_{cor})]
\end{equation}
where $\sigma(t,\tau_{cor})$ is a uniform random function lying in the range [+1,-1], changing values at a coherence time, $\tau_{cor}$. Statistical fluctuations are characterized by $\delta$ and $\tau_{cor}$, which correspond to percentile level of fluctuation and coherence time correspondingly. Figure \ref{fig2} shows a typical $\alpha $ fluctuation generated by our random number generation program. Stochastic variations in the Babcock-Leighton $\alpha$ coefficient are natural because they arise from the cumulative effect of a finite number of discrete flux emergences, i.e., active region eruptions, all with various degrees of tilt randomly scattered  around a mean Joy's law distribution.

In this system the dynamo number ($N_D= \alpha_0 \omega \tau^2/L$) is defined as the ratio between the source and dissipative terms, which is a measure of the efficiency of the dynamo mechanism. The product of source terms is $|\alpha_0 \omega/L|$ while that of the dissipative terms is $1/\tau^2$. In terms of physical parameters, the expected diffusion time scale ($L^2/\eta$) in the SCZ is 13.8 yr for a typical diffusivity of $10^{12} cm^2 s^{-1}$ implying that the dissipative term ($1/\tau^2$) is of the order of $10^{-18} s^{-2}$. Now, if we take the value of $\omega$ as the difference in rotation rate across the SCZ in nHz (as measured; for details see \cite{howe09}) , L as the length of SCZ and $\alpha_0$ as 1 m $s^{-1}$ then the source term $|\alpha_0 \omega/L|$, is of the same order as the dissipative term and the dynamo number can be made higher than unity by slightly adjusting the $\alpha$ coefficient. In fact, if the tachocline is considered as the interface across which flux transport is occuring, then the dynamo number becomes even greater as the radial differential rotation is about the same while the length scale reduces further. In this model we always take the value of $|\alpha_0 \omega/L|$ (source term) to be greater than $1/\tau^2$ (decay term), and set the magnitude of $|\omega/L|$ and $|\alpha_0|$ in a way such that the strength of toroidal field is greater than the strength of poloidal field (as suggested by observations). In summary, keeping all of the other physically motivated parameters fixed, the dynamo number can be varied by adjusting the value of $\alpha_0$. Since $B_{min}$ corresponds to the equipartion field strength (on the order of $10^4$ Gauss) above which magnetic flux tubes become buoyant while $B_{max}$ is on the order of $10^5$ Gauss (above which flux tubes emerge without any tilt, thus shutting off the Babcock-Leighton source; \cite{silva93}), we take the ratio of $B_{max}/B_{min}$ as 7 for all of our calculation. Here we explore our low order time delay model in two parameter space regimes to test for robustness. In the first case we fix the parameters as $ \tau=15, B_{min}=1, B_{max}=7, T_0=4 T_1, T_1=0.5$ and $\omega/L=-0.34$ while in the second case we take $ \tau=25, B_{min}=1, B_{max}=7, T_0=40 T_1, T_1=0.5$ and $\omega/L=-0.102$.  Initial conditions are taken to be $(B_{min}+B_{max})/2$ for both A and $B_\phi$.  Our choice of parameters ensures that in both cases the diffusive timescale is much higher than flux transport timescales ($\tau > T_0 + T_1$). The simulations are robust over a range of negative $N_D$ values; for a detailed parameter space study of the underlying model without stochastic fluctuations, please refer to \cite{Wilmot-Smith06}. Below, we present the results of our stochastically forced dynamo simulations focussing on entry and exit from grand minima episodes.

\section{Results and Discussions}

We first perform simulations without the lower operating threshold in the Babcock-Leighton $\alpha$-effect (setting ${B_{min}} = 0$) in Eqn.~3. A majority of Babcock-Leighton dynamo models, including many that have explored the dynamics of grand minima do not use this lower operating threshold. As already known \citep{Charbonneau2000,Choudhuri2012} we find that the Babcock-Leighton dynamo with this setup generate cycles of varying amplitudes, including episodes of higher than average activity levels (grand maxima) and occasional episodes of very low amplitude cycles reminiscent of Maunder-like grand minima (Fig.~3, upper panel; Fig.~4, upper panel). When we do switch on the lower operating threshold, however, we find that the Babcock-Leighton dynamo is unable to recover once it settles into a grand minimum
(Fig.~3, lower panel; Fig.~4, lower panel). This striking result can be explained invoking the underlying physics of the solar cycle. When a series of poloidal field fluctuations lead to a decline in the toroidal field amplitude below the threshold necessary for magnetic buoyancy to operate (with a consequent failure of sunspots to form), the Babcock-Leighton poloidal field source which relies on bipolar sunspot eruptions completely switches off resulting in a catastrophic quenching of the solar cycle. Earlier simulations, which did not include the lower quenching missed out on this physics because even very weak magnetic fields, which in reality could never have produced sunspots, continued to (unphysically) contribute to poloidal field creation. Earlier, it has been shown that the lower threshold due to magnetic buoyancy plays a crucial amplitude limiting role in the Babcock-Leighton solar cycle \citep{Nandy02} and this study indicates that this should be accounted for in all Babcock-Leighton solar dynamo models.

\begin{figure}[t!]
\centering
\includegraphics*[width=\linewidth]{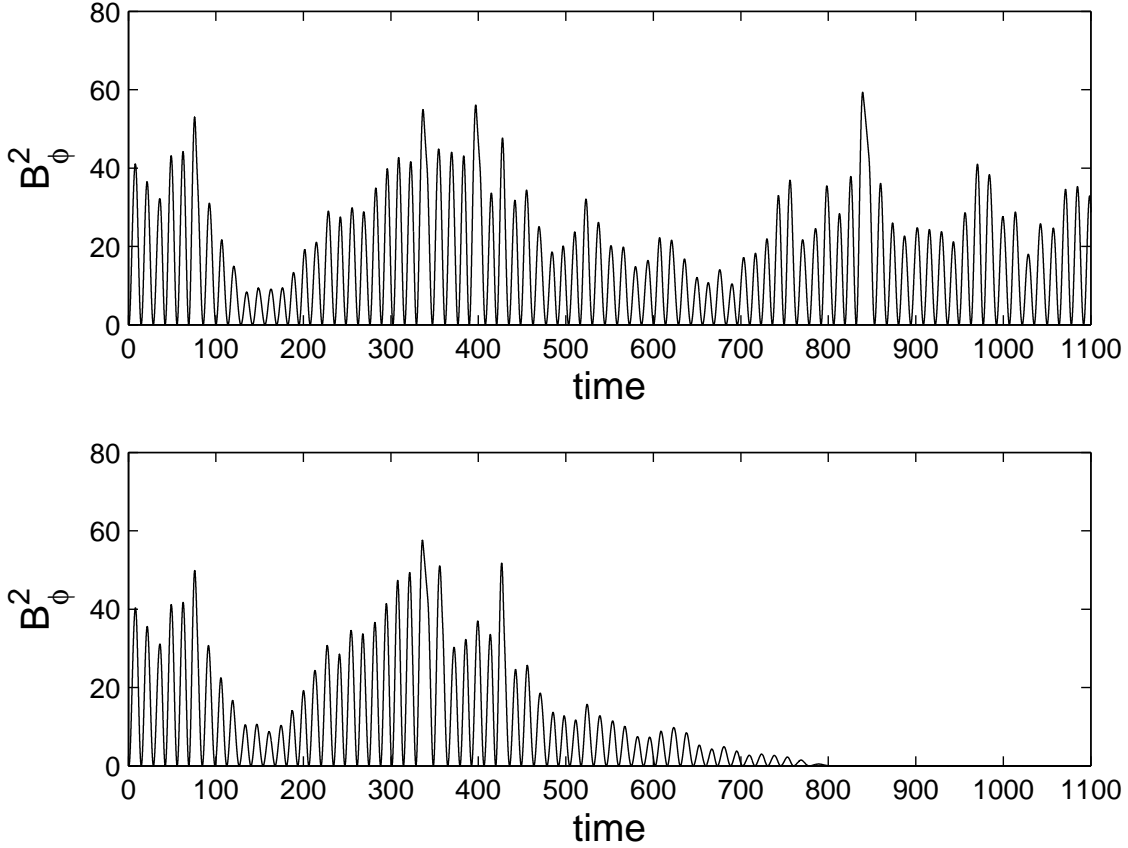}
  \caption{(a) Time evolution of the magnetic energy proxy without considering the lower operating threshold in the quenching function ($B_{min}=0$); (b) Same as above but with a finite lower operating threshold ($B_{min}=1$). The solar dynamo never recovers in the latter case once it settles into a grand minima. All other parameters are fixed at $ \tau=15, B_{max}=7, T_0=2, T_1=0.5, \omega/L=-0.34$ and $\alpha_0=0.17$}
\label{fig3}
\end{figure}

\begin{figure}[t!]
\centering
\includegraphics*[width=\linewidth]{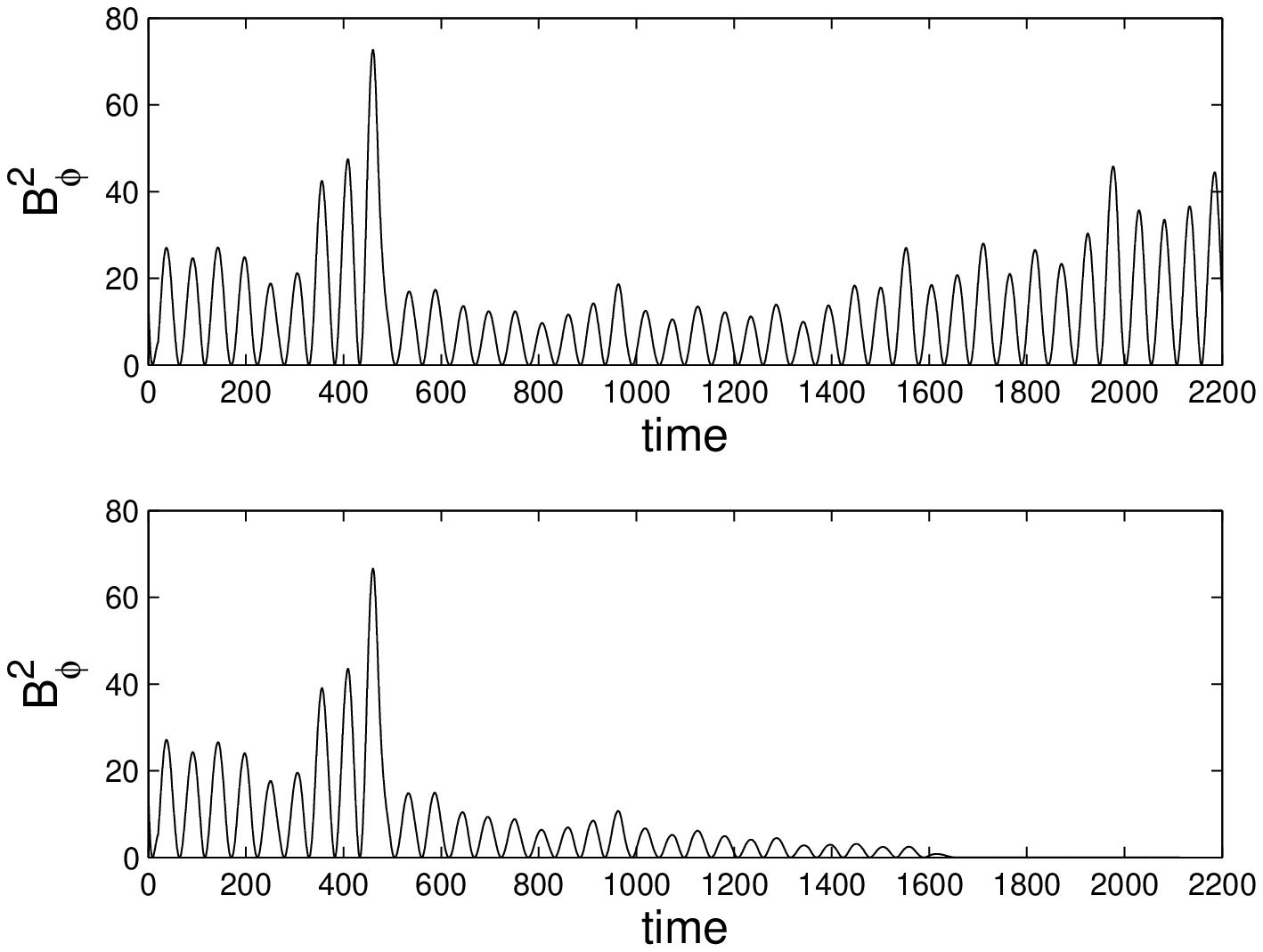}
  \caption{(a) Time evolution of the magnetic energy proxy without considering the lower operating threshold in the quenching function ($B_{min}=0$); (b) Same as above but with a finite lower operating threshold ($B_{min}=1$). The solar dynamo never recovers in the latter case once it settles into a grand minima. All other parameters are fixed at $ \tau=25, B_{max}=7, T_0=20, T_1=0.5, \omega/L=-0.102$ and $\alpha_0=0.051$}
\label{fig3}
\end{figure}
To circumvent this problem faced by the stochastically forced Babcock-Leighton dynamo, we explicitly test an idea \citep{Nandy2012} for the recovery of the solar cycle based on an additional poloidal source effective on weak toroidal fields. Since the tachocline is the seat of strong toroidal field, any weak field $\alpha$ which is effective only on sub-equipartition strength field will get quenched there. Thus, this $\alpha$-effect must reside above the base of the SCZ \citep{Parker1993} in a layer away from the strongest toroidal fields. Motivated by this, we devise a new system of dynamo equations governed by
\begin{eqnarray}
    \frac{dB_\phi(t)}{dt} &=& \frac{\omega}{L} A(t-T_0)-\frac{B_\phi(t)}{\tau}\\
   \frac{dA(t)}{dt} &=& \alpha_0 f_1(B_\phi(t-T_1)) B_\phi(t-T_1) \nonumber \\
    &+& \alpha_{mf} f_2(B_\phi(t-T_2)) B_\phi(t-T_2) -\frac{A(t)}{\tau} \nonumber \\
\end{eqnarray}
where $f_2$, the quenching function for the weak field poloidal source $\alpha_{mf}$ is shown in Fig.~1 and is parameterized by
\begin{equation}
f_2 = \frac{erfc(B^2_\phi(t - T_2)-B^2_{eq})}{2}.
\end{equation}
Taking $B_{eq}=1$ ensures that the weak field source term gets quenched at or below the lower operating threshold for the Babcock-Leighton $\alpha$ and the former, therefore, can be interpreted to be the mean field $\alpha$ effect. In equation~6, the time delay $T_2$ is the time necessary for the toroidal field to enter the source region where the additional, weak-field $\alpha$ effect is located.

If $T_2=0$, i.e. the generation layer of the additional $\alpha$ effect is coincident with the $\Omega$ effect (layer) then we find that the stochastically forced dynamo again fails to recover from a grand minimum. This is reminiscent of the original motivation behind the introduction of the interface dynamo idea with spatially segregated source regions \citep{Parker1993}. However, if $T_2$ is finite and $T_1 > T_2$ (i.e., there is some segregation between the $\Omega$ effect toroidal source, the additional weak-field $\alpha$ and the Babcock-Leighton $\alpha$), we find that the solar cycle can recover from grand minima like episodes in a robust manner. Figures 5 and 6 depict such solutions (for two different sets of parameter), where we explicitly demonstrate self-consistent entry and exit from grand minima like episodes. We note that this recovery of the solar cycle from grand minima like episodes is possible with or without fluctuations in the additional, weak field poloidal source term $\alpha_{mf}$.

\begin{figure}[t!]
\centering
\includegraphics*[width=\linewidth]{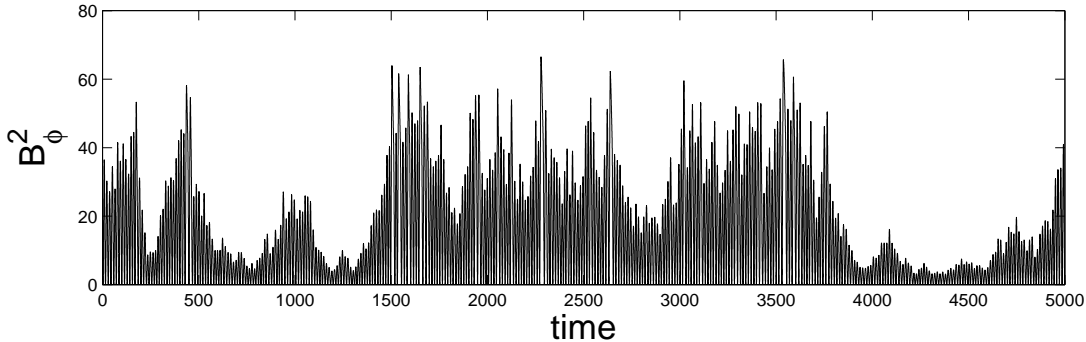}
  \caption{Time series of the magnetic energy ($B_\phi^2$) with both Babcock-Leighton and a weak (mean-field like) $\alpha$ effect for 30\% fluctuation in $\alpha$, $\tau$=15, $T_0$=2, $T_1$=0.5, $T_2$=0.25, $B_{min}$ = $B_{eq}$ =1, $B_{max}$=7, $\omega/L$=$-0.34$, $\alpha_0$=0.17 and $\alpha_{mf}$=0.20. This long-term simulation depicts the model's ability to recover from grand minima episodes.}
\label{fig4}
\end{figure}
\begin{figure}[t!]
\centering
\includegraphics*[width=\linewidth]{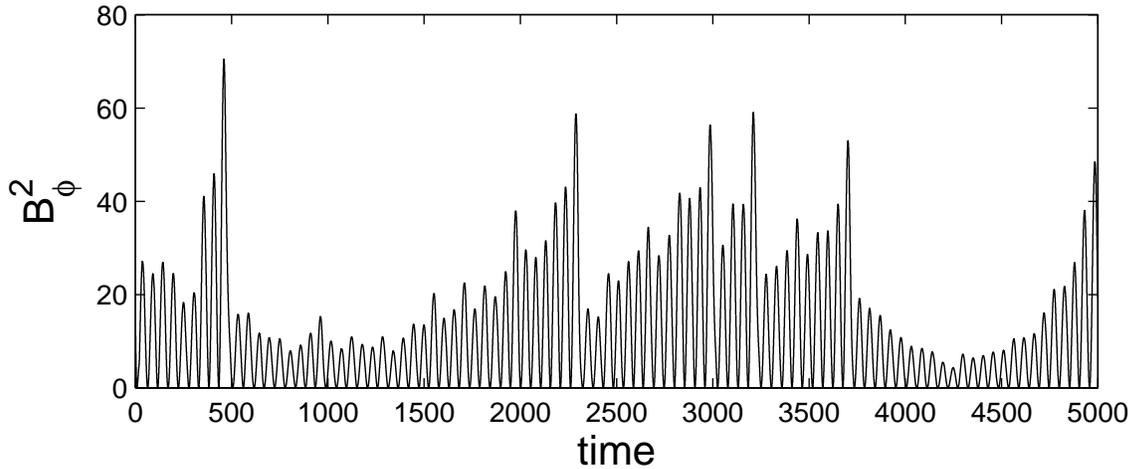}
  \caption{Time series of the magnetic energy ($B_\phi^2$) with both Babcock-Leighton and a weak (mean-field like) $\alpha$ effect for 50\% fluctuation in $\alpha$, $\tau$=25, $T_0$=20, $T_1$=0.5, $T_2$=0.25, $B_{min}$ = $B_{eq}$ =1, $B_{max}$=7, $\omega/L$=$-0.102$, $\alpha_0$=0.051 and $\alpha_{mf}$=0.04. This long-term simulation depicts the model's ability to recover from grand minima episodes.}
\label{fig4}
\end{figure}

\section{Conclusions}

In summary, we have constructed a new model of the solar dynamo for exploring solar cycle fluctuations based on a system of stochastically forced, non-linear, delay differential equations. Utilizing this model for long-term simulations we have explicitly demonstrated that the currently favored mechanism for solar poloidal field production, the Babcock-Leighton mechanism, alone, cannot recover the solar cycle from a grand minimum. We have also demonstrated that an additional, mean field like $\alpha$-effect capable of working on weaker fields is necessary for self-consistent entry and exit of the solar cycle from grand minima episodes. We have demonstrated that our results and conclusions hold over two very diverse regimes of parameter choices. We note that simulations motivated from this current study and based on a spatially extended dynamo model in a solar-like geometry supports the results from this mathematical time delay model \citep{Passos2014}. Taken together, these strengthen the conclusion that a mean field like $\alpha$-effect effective on weak toroidal fields must be functional in the Sun's convection zone and that this is vitally important for the solar cycle, even if the dominant contribution to the poloidal field comes from the Babcock-Leighton mechanism during normal activity phases.

\acknowledgements{We are grateful to the Ministry of Human Resource Development, Council for Scientific and Industrial Research, University Grants Commission and the Ramanujan Fellowship award  of the Department of Science and Technology of the Government of India for supporting this research. D.P. acknowledges support from the Funda\c{c}\~{a}o para a Ci\^{e}ncia e Tecnologia grant SFRH/BPD/68409/2010, CENTRA and the University of the Algarve and is also grateful to IISER Kolkata for hosting him during the performance of this research.}

\end{document}